# Solar One: A Proposal for The First Crewed Interstellar Spacecraft

Alberto Caballero

**Abstract:** In this paper it is presented the concept and design of a new type of spacecraft that could be used to make the first manned interstellar travel. Solar one would integrate three near-term technologies, namely: compact nuclear fusion reactors, extremely large light sails, and high-energy laser arrays. A system of lenses or mirrors to propel the sail with sunlight is suggested as an alternative to laser propulsion. With a mile-long light sail, Solar One could reach an average of 22% the speed of light, arriving to the closest potentially habitable exoplanet in less than 19 years with the help of a Bussard scoop producing reverse electromagnetic propulsion. Key challenges are reducing the weight of continuous-wave lasers and compact fusion reactors as well as achieving cryo-sleep and artificial gravity.

## 1  Introduction

Several light-sail spacecrafts have already been tested. Some examples are LightSail 1 and LightSail 2, from the Planetary Society (2020). However, these light sails are propelled by sunlight, and the solar radiation pressure is very small (just 6.7 Newtons per gigawatt, which equals to 9 Newtons/km$^2$ or 1,400 watts/m$^2$ at 1 AU).

Lasers can provide radiation pressures much higher than the Sun, unless Fresnel lenses are used. Previous experiments with directed-energy weapons have proved successful. The Boeing YAL-1, a spacecraft equipped with a Kilowatt-laser, was able to deliver a power density over 100 watts/cm$^2$ at a distance of 1 km (US Air Power, 2008).

In 2016, scientists announced the first design for a beamed-powered spacecraft that could reach speeds of 0.2c. The project, called StarShot, entailed the idea of sending 1,000 nanocrafts with light sails attached that would be powered by a 100 GW laser array (Breakthrough Initiatives, 2020). As of today, the concept is still considered to be the best option for unmanned interstellar travel. Potentially habitable exoplanets such as Proxima b could be reached in only 20 years.

However, the idea is to send the Starshot nanocrafts 1 AU away from Proxima b. This distance would be enough to photograph the exoplanet, but perhaps not sufficient to notice the presence of a possible intelligent civilization less advanced than humanity. For this reason, and to better study the exoplanet, a manned interstellar spaceship becomes necessary.

The proposal more similar to Solar One was presented by Robert Forward in 1984. Forward proposed a 64-tonne and 30-km payload sail surrounded by a 644-tonne and 100-km decelerator sail, launched by a 7.2-Terawatt laser system to reach 21 percent the speed of light. As the spacecraft approaches Alpha Centauri, another 26-TW laser is pointed towards a 1000-km Fresnel lens that decelerates the spacecraft. Another similar proposal is the Laser Powered Interstellar Rocket (Jackson, 1978). In this case, a laser array beams to a collector on the spacecraft, without the use of a light sail.

## 2 Concept

Solar One is a design for a manned spaceship that would be powered by beamed and photon propulsion. The name 'Solar One' has been chosen to better represent our civilization: the term 'Solar' refers to the solar system and the term 'One' refers to the first design of a possible fleet of future spaceships. Three are the technologies that would be used: the US Navy CFR nuclear fusion reactor, a slightly larger version of the NASA's Sunjammer light sail, and a laser system called DE-STAR (Directed Energy System for Targeting of Asteroids and exploRation).

Firstly, the US Navy CFR (Compact Fusion Reactor) is a mobile unit that could provide one Terawatt of power (Forbes, 2019). Instead of using large superconducting magnets, this reactor uses small conical dynamic fusors (Salvatore Pais, 2018).

Solar One would carry a 1-Terawatt reactor on board to power the laser system needed for deceleration. The system would have to incorporate radiators to dissipate the heat. Lockheed Martin has already suggested the possibility of building 200-ton reactors of 200 Megawatts (Besa Center, 2018). In other models, a 'blanket' component would transfer energy to a coolant, making the whole reactor to weight as low as 300 tons (The Enginer, 2014). Although this would be far from what Solar One needs, it would be a starting point.

Antimatter would be a better source of energy for the laser. Only a few milligrams of antimatter could yield power densities in the Terawatt order. To produce the antimatter needed, an engine such as the VARIES Mk 1 could generate Schwinger antiparticle pairs directly from the vacuum (Next Big Future, 2012). The problem with this resides in the fact that a laser in the Petawatt order would be needed. The matter-antimatter GeV gamma ray laser proposed by Winterberg in 2012 could also be used to decelerate Solar One, but creating the antimatter needed would be difficult with current technology.

Secondly, the Sunjammer light sail is a proposed NASA sail with a size of 38 x 38 m (1,444 square meters) (NASA, 2017). Solar One would need a 1 mile-light sail (1,609 by 1609 meters). A robust structure would be needed to support such light sail.

Thirdly, DE-STAR is a 100-GW proposed laser intended for asteroid destruction and beamed-power propulsion. Solar One would need 260 of these lasers in order to achieve the power output suggested by Forward (1984) of 26 TW. The laser system could be placed either on Earth or a Lagrange point. Current DE-STAR proposals would be useless for Solar One, but more powerful and lighter versions of the system could be used.

To accelerate the spacecraft, an alternative solution would be to use Fresnel lenses 85 meters wide, which could provide around 10 MW/m$^2$. A system of flexible mirrors would be needed to continuously focus the light into the sail. The problem is that more than 2.5 million lenses would be needed to propel a one-mile sail. Therefore, it would probably be easier to build just one lens of around 137 km able to provide 26 TW of power, or several lenses of an intermediate size.

Another alternative would be to build and place on Earth a large parabolic mirror able to gather those 26 TW of power needed. In order to obtain such energy, a 137-km mirror would be needed. The problem in this case is that at least two primary mirrors would be needed, one on each side of the Earth. The light would be sent to a flexible secondary mirror placed in orbit or a Lagrange point.

Whether a Fresnel lens or parabolic mirror are used, a size of 137 km would not be feasible. It would be necessary to send a Fresnel lens closer to the Sun. Considering that the solar irradiance in Mercury is 9,116 watts/m$^2$, a Fresnel mirror placed in its orbit would have to be 54 km wide, which still might not be feasible. However, at the distance where the Parker Solar Probe will get its limit of 650 Kw/m$^2$ (IEEE Spectrum, 2019), a 6.4-km lens or mirror could gather 26 TW of power. This would give Solar One an acceleration of 0.06g. For 1g, an estimated mirror array of 100 km would be necessary, which could be divided into for example 10 mirrors of 10 km each or 100 mirrors of 1km each.

As it was mentioned before, Solar One would have a 1-TW laser system at the front of the cockpit in order to decelerate at destination using photon propulsion. The spacecraft would need 10 DE-STAR lasers in order to achieve the required power output. A similar project called HELLADS (High Energy Liquid Laser Area Defense System) has a target mass of 5 tons for a 1 MW-laser (5 kg per kW) (Air Force Technology, 2018). Although this is, again, far from what Solar One needs, it could be a starting point.

Carrying the necessary nuclear fuel on board might not be the best option. As in the Bussard Ramjet, it would be possible to collect hydrogen from space with a scoop placed at the front of the cockpit. The electromagnetic fields produced would drastically reduce the time of deceleration.

To accelerate Solar One, if we use the power output of 26 TW suggested by Forward (1984), a 500-mm laser placed for example 1.5 million meters away from Earth and with, for example, a divergence of 0.07° would produce an irradiance of more than 10 MW/m². These calculations were made with an online calculator provided by Kvant lasers (2020), a company specialized in lasers.

| | |
|---|---|
| Beam diameter at aperture: | 500 mm |
| Divergence: | 1.21 mrad |
| Distance to audience: | 1500000 m |
| Laser power: | 260000 mW |
| Diameter at audience: | 1815500.0 mm |
| Minimum diameter (> 7mm): | 1815500.0 mm |
| Beam area: | 2588703958834.1 mm² |
| Irradiance: | 1004363.6 mW/cm² |

Figure 1: Power density at the beginning of the acceleration Source: Kvant Lasers

Due to the square inverse law, as the spacecraft moves away from the laser, the power density reduces. To maintain the same power density all the time, it would be necessary to incorporate in the laser system an automatic parabolic mirror that would gradually increase its diameter in order to reduce the divergence of the beam to a value close to zero.

The mirror could be formed by several small mirrors, each of them with a specific orientation every time. An alternative would be to place the mirrors outside the laser system, or even in space.

| | |
|---|---|
| Beam diameter at aperture: | 500 mm |
| Divergence: | 0.00000 mrad |
| Distance to audience: | 283629' m |
| Laser power: | 260000 mW |
| Diameter at audience: | 1815726.7 mm |
| Minimum diameter (> 7mm): | 1815726.7 mm |
| Beam area: | 2589350595382.6 mm² |
| Irradiance: | 1004112.8 mW/cm² |

Figure 2: Power density at the end of the acceleration Source: Kvant Lasers

The idea behind Solar One is to combine the three projects. A 4-crew spaceship with a target mass of 300 tons could be powered by a mile-long light sail and achieve the speed of 0.3c with a constant acceleration during the first 5 years of the trip.

The major breakthrough would have to come for reducing the mass of both the nuclear reactor and laser. Portable Terawatt pulsed laser systems such as Teramobile have already been built (Centre National de la Recherche Scientifique, 2008), but Solar One would need a continuous-wave laser.

No light sail is able to reflect 100 percent of the light. A carbon fibre sail with a reflectivity of 90 percent and able to withstand 2,770 K could be used. Such sail would have a density of 2.65 g/cc, a thickness of 1 micron, and a total mass of 10.7 tons.[2] However, to simplify the math, the following calculations are made considering a 100% sail efficiency.

The resulting force is calculated with the following equation:

$$F = 2(P \times A)/C$$
$$F = 2(10,043,630 \times 2,588,881)/300,000,000$$
$$F = 173,345 \text{ newtons}$$

F* = force / thrust (newtons)
P* = power (watts /m$^2$)
A* = surface area of light sail (m$^2$)
c* = speed of light

The acceleration obtained would be 0.57 m/s$^2$:

$$a = F/M$$
$$a = 173,345/300,000$$
$$a = 0.57 \text{m/s}^2$$

a* = acceleration (m/s$^2$)
M* = mass (kg)

And it would take almost 5 years to reach 0.3 c (89 million m/s):

$$t = v/a$$
$$t = 89,940,000/0.57$$
$$t = 155,659,397 \text{sec} = 4,9 \text{years}$$

---

[2] Calculations made with this Solar Sail Calculator: http://www.georgedishman.f2s.com/solar/Calculator.html.

For the deceleration, considering that photons have a thrust to power ratio of 3.34 x 10$^{-9}$ newtons per watt, the 1-TW laser system would provide the following opposite force:

$$F = 1 \times 10^{12} \text{ watts } \times 3.34 \times 10^{-9} = 3,340 \text{N}$$

To use the power and deceleration time just mentioned, the spacecraft would have to detach from the light sail, ideally just after the acceleration stage finishes. The deceleration with the laser and without light sail would be 0.011 m/s$^2$:

$$a = F/M$$
$$a = 3,340/289,300$$
$$a = 0.011 m/s^2$$

Considering that a proton-proton fusion reaction can produce up to 645 TW, 3.8 grams of hydrogen per second would provide a thrust similar to during the acceleration. The density of hydrogen atoms is about one H-atom per cubic centimeter. For the scoop to generate the electromagnetic fields, an annular copper cylinder coated with a layer of superconducting tin-niobium alloy (Nb3Sn) has already been proposed. However, considering that a 1,000-km scoop would weight around 200 tons and collect 60 grams of hydrogen per second at 0.15c, to collect 3.8 grams per second the scoop would have a size of 63 km, which is not feasible.

Moreover, using the photon rocket on-board to ionize the hydrogen would increase the amount of fuel scooped and reduce the size of the scoop. The CNO cycle would reduce the fusion reactor size to at least 10 times in dimension (Centauri Dreams, 2020). The electromagnetic fields would help to decelerate the spacecraft. The average drag is calculated with the following equation (Zubrin, 1990):

$$d = m \times v$$
$$d = 0,0038 \times 44,969,000$$
$$d = 170,882 \text{N}$$

d* = drag (N)
m* = mass of collected hydrogen (kg/s)
v* = average speed during deceleration (m/s)

This force together with the force produced by the photon rocket would be 174,222 newtons. The resulting acceleration would be 0.6 m/s$^2$, and it would take 4,75 years to stop a spacecraft travelling at 0.3c.

$$t = v/a$$
$$t = 89,940,000/0.6$$
$$t = 149,940,000 \text{sec} = 4,75 \text{ years}$$

At an average speed of slightly more than 0.22 c, the crew would arrive to the Alpha Centauri system in slightly less than 19 years. The spacecraft would be either accelerating or decelerating during half of the travel.

$$\text{Average speed} = 0.51 \cdot 0.15 +$$
$$0.49 \cdot 0.3 = 0.2235c$$

0.51* = percentage of trip time during acceleration and deceleration
0.15* = average speed during acceleration and deceleration
0.49* = percentage of trip time with constant speed
0.3* = cruise speed

Duration of the trip to Alpha Cen = Distance / Average speed = 4.24/0.2235 ≈ 18.9 years

## 3  Design

The spaceship would be composed of the following main elements: a laser system, a light sail, a nuclear micro-reactor, a Bussard scoop, and a cockpit that could also be used as a descent module. An extra amount of light sail would be ideal in case of damage caused by micro-asteroids.

Detaching the sail after the acceleration stage would be ideal to reduce weight for the deceleration. The spacecraft could be located behind the light sail, as in the following image.

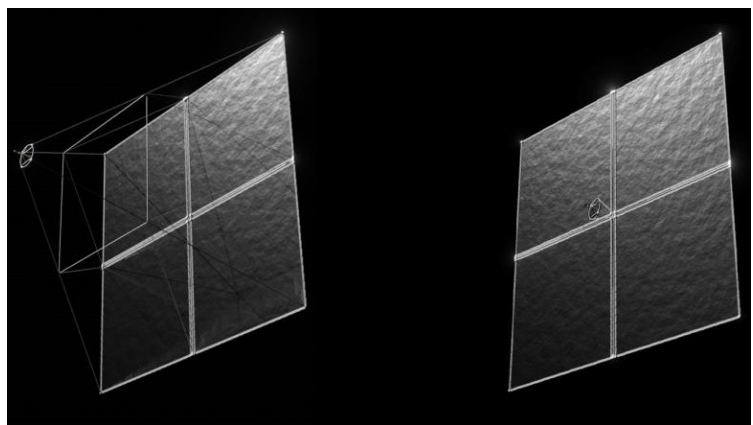

Figure 3: Solar One
Illustrations made by Marco Purich

During the acceleration, the light of the laser reflected by a primary mirror placed inside the laser system would be sent to a secondary parabolic or hyperbolic mirror that would also gradually change its form in order to give the divergence needed at any given moment.

Instead of using fixed mirrors, they could gradually change their orientation in order to change the divergence of the beam. However, as it was mentioned before, these mirrors could be placed outside the laser system.

The light sail could be placed in the middle of the sail instead of behind, reducing the amount of structure needed and probably the overall weight of the spacecraft (see figure 4).

As with the spacecraft collecting fuel from space, the ideal would be to use sunlight to propel the light sail. Several systems composed by one Fresnel lens and two parabolic mirrors could be sent to different places of the Solar System. As they rotate around the Sun, each of them would be aimed at the light sail of Solar One during a specific time. These mirrors would also have to gradually change their orientation to continuously keep the laser beam focused.

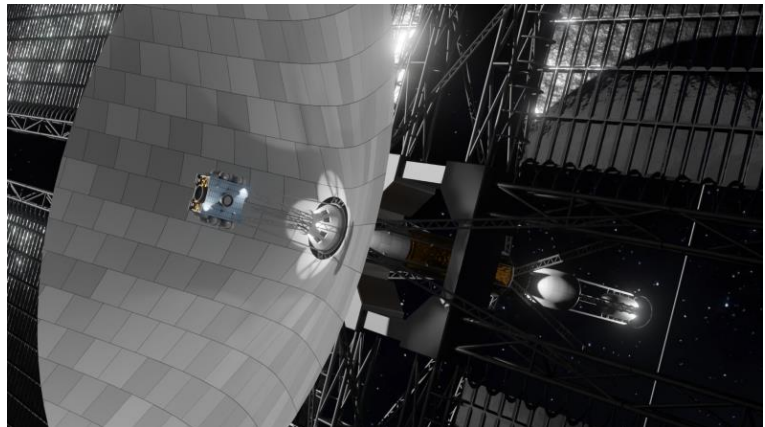

Figure 4: Solar One - From the right: nuclear reactor, cockpit, Bussard scoop, laser system
Illustration made by Marco Purich

Once the destination is reached, the crew could orbit the exoplanet, take images and send a robot to the surface. If the conditions are safe, the crew could choose to land in order to personally explore the exoplanet.

## 4  Challenges

Engineers would likely face several challenges while building Solar One. One of them would be to build a large array of DE-STAR lasers and, most importantly, produce the necessary energy to power them during five years (Hughes, 2013). Another obstacle would be to reduce the weight of the on-board nuclear reactor and laser system. Other challenges would be to protect all the modules from dust and micro- asteroid impacts, to protect the cockpit from cosmic radiation, to reduce the amount of energy and supplies needed by the crew, and to reduce the effects of zero gravity.

All the modules would have a special shielding. Cryo-sleep would be the most efficient way to make a 19-year trip. In this line, NASA has been working on a cryo-sleep design called TORPOR (Torpor Inducing Transfer Habitat For Human Stasis To Mars) (NASA, 2013). Lastly, rotational simulated gravity has been proposed as a solution to zero gravity.

# 5 Conclusions

In this paper it has been analysed the possibility of building a manned interstellar spaceship with a light sail propelled by an external laser system, parabolic mirrors, or lenses. To decelerate, the spacecraft would use an on-board laser system that would receive the necessary electricity from a small nuclear fusion reactor. This reactor would obtain fuel from a Bussard scoop which, in turn, would help to decelerate the spacecraft.

Small modular reactors such as the US Navy CFR have already been patented, large light sails such as Sunjammer have already been built, and directed energy weapons such as DE-STAR have been proposed for interstellar travel. Nuclear fusion is the most near-term technology that could be used to power Solar One.

However, research on antimatter propulsion is advancing at a rapid rate. If scientists are able to produce more antimatter than the energy used to generate it, this would be the best way to power Solar One.